\documentclass[useAMS,usenatbib]{mn2e} 
\usepackage{aas_macros}
\usepackage{graphics}
\usepackage[pdftex]{graphicx}
\usepackage{epstopdf}
\usepackage{epsfig}  
\usepackage{natbib} 
\usepackage{float}
\usepackage{amsmath}
\usepackage{times}
\usepackage[varg]{txfonts}
\usepackage{verbatim} % need that for multi-line comments
\usepackage{multirow,bigdelim} % to do the large brackets in the mock table
\usepackage{color}

\newcommand{\icecore}{{\sc ICeCoRe}}

\newcommand{\bl}{\boldsymbol}
\newcommand{\tm}{\textrm}

\newcommand{\hMpc}{{\ifmmode{h^{-1}{\rm Mpc}}\else{$h^{-1}$Mpc}\fi}}
\newcommand{\hkpc}{{\ifmmode{h^{-1}{\rm kpc}}\else{$h^{-1}$kpc}\fi}}
\newcommand{\hMsun}{{\ifmmode{h^{-1}{\rm {M_{\odot}}}}\else{$h^{-1}{\rm{M_{\odot}}}$}\fi}}

\def\lesssim{\mathrel{\hbox{\rlap{\hbox{\lower4pt\hbox{$\sim$}}}\hbox{$<$}}}}
\def\gtrsim{\mathrel{\hbox{\rlap{\hbox{\lower4pt\hbox{$\sim$}}}\hbox{$>$}}}}
\title[Initial conditions from peculiar velocities]
      {Reconstructing cosmological initial conditions from galaxy peculiar velocities. II. The effect of observational errors}
\author[Doumler et al.] 
{Timur Doumler$^{1,2}$, H\'el\`ene Courtois$^{1}$, Stefan Gottl\"ober$^{2}$, and Yehuda Hoffman$^{3}$  \\
  $^1$Universit\'e Lyon 1, CNRS/IN2P3, Institut de Physique Nucl\'eaire, 69622 Villeurbanne, Lyon, France\\
  $^2$Leibniz-Institut f\"ur Astrophysik Potsdam, An der Sternwarte 16, 14482 Potsdam, Germany\\
  $^3$Racah Institute of Physics, Hebrew University, Jerusalem 91904, Israel\\
  }

\begin{document}

\date{}

\pagerange{\pageref{firstpage}--\pageref{lastpage}} \pubyear{2012}

\maketitle

\label{firstpage}

%%%%%%%%%%%%%%%%%%%%%%%%%%%%%%%%%%%%%%%%%%%%%%%%%%%%%%%%%%%%%%%%%%%%%%%%%%%

\begin{abstract}

The Reverse Zeldovich Approximation (RZA) is a reconstruction method which allows to estimate the cosmic displacement field from galaxy peculiar velocity data and to constrain initial conditions for cosmological simulations of the Local Universe. In this paper, we investigate the effect of different observational errors on the reconstruction quality of this method. For this, we build a set of mock catalogues from a cosmological simulation, varying different error sources like the galaxy distance measurement error (0 -- 20\%), the sparseness of the data points, and the maximum catalogue radius (3000 -- 6000 km/s). We perform the RZA reconstruction of the initial conditions on these mock catalogues and compare with the actual initial conditions of the simulation.  We also investigate the impact of the fact that only the radial part of the peculiar velocity is observationally accessible. We find that the sparseness of a dataset has the highest detrimental effect on RZA reconstruction quality. Observational distance errors also have a significant influence, but it is possible to compensate this relatively well with Wiener Filter reconstruction. We also investigate the effect of different object selection criteria and find that distance catalogues distributed randomly and homogeneously across the sky (such as spiral galaxies selected for the Tully-Fisher method) allow for a higher reconstruction quality than if when data is preferentially drawn from massive objects or dense environments (such as elliptical galaxies). We find that the error of estimating the initial conditions with RZA is always dominated by the inherent non-linearity of data observed at $z=0$ rather than by the combined effect of the observational errors. Even an extremely sparse dataset with high observational errors still leads to a good reconstruction of the initial conditions on a scale of $\approx 5$ Mpc/$h$.

\end{abstract}
%%%%%%%%%%%%%%%%%%%%%%%%%%%%%%%%%%%%%%%%%%%%%%%%%%%%%%%%%%%%%%%%%%%%%%%%%%%

\begin{keywords}
  cosmology: theory -- dark matter -- large-scale structure of Universe --
  galaxies: haloes -- methods: numerical
\end{keywords}

%%%%%%%%%%%%%%%%%%%%%%%%%%%%%%%%%%%%%%%%%%%%%%%%%%%%%%%%%%%%%%%%%%%%%%%%%%%
\section{Introduction}
\label{sec:introduction}

During the last years there has been an impressive advancement in the field of modelling the formation and evolution of the large-scale structure (LSS) of the Universe. In large part, our understanding of this highly non-linear process is enabled by conducting cosmological $N$-body simulations, which usually model the time evolution of a finite subvolume of the Universe from the primordial density perturbations through the LSS formation until the present epoch. These methods complement observations of our Universe at large scales, such as the properties and distributions of galaxies. Here, the Local Universe, our cosmological neighbourhood, is the best-studied region. 

A very attractive approach to link together these observed local structures and the complementary simulation techniques is the constrained realizations (CR) method. With the CR algorithm \citep{Hoffman1991}, it is possible to constrain the initial conditions (ICs) of cosmological simulations using as input observational data of the Local Universe. The resulting constrained simulations are able to reproduce the major objects of the Local Universe: the Local Supercluster (LSC) with the Virgo cluster, the Great Attractor (GA), the Local Void, and the Coma and Perseus-Pisces clusters. The CLUES project\footnote{www.clues-project.org} is an international collaboration of theoretical and observational cosmologists with the goal of producing such simulations with the highest accuracy possible \citep{Gottloeber2010arXiv}. 

A crucial part of research within the CLUES project and the motivation for this work is the question of how constrained ICs are best generated. Initially, galaxy redshift catalogues were used as the input data for constrained simulations \citep{Kolatt1996,Bistolas1998,Mathis2002,Lavaux2010b}. However, galaxy peculiar velocity measurements provide a valuable alternative to redshift data for generating constrained ICs \citep{Klypin2003,Gottloeber2010arXiv}. The radial component of galaxy peculiar velocities can be derived from their redshift and an independent measurement of their luminosity or distance. Peculiar velocities provide a direct tracer of the total mass distribution (without galaxy bias), and the associated distances provide galaxy positions in real space (without redshift distortions). Further, galaxy peculiar velocities are strongly correlated over large distances, and therefore allow for a reconstruction of the underlying field over large distances and despite the sparse and inhomogeneous sampling. For the same reason, they are less sensible to nonlinear effects of structure formation, which become increasingly important at smaller scales. The theoretical framework necessary to use radial peculiar velocities as constraints for ICs was developed by \citet{Zaroubi1999}, combining the theory of Gaussian random fields and their reconstruction from sparse and noisy data sets using a Bayesian approach with the CR algorithm.

With a measurement of the distance $r$ to a galaxy with an estimated measurement error $\Delta r$ ,and the observed redshift velocity $v_r^{\tm{obs}}$, the radial peculiar velocity $v_r^{\tm{pec}}$ and its measurement error $\Delta v_r^{\tm{pec}}$ are obtained via
\begin{align}
v_r^{\tm{pec}} &= v_r^{\tm{obs}} - r \cdot H_0\;\;\;, \\
\Delta v_r^{\tm{pec}} &= - \Delta r \cdot H_0 \;\;\;\;\;\;\,\,.
\end{align}
Several methods exist to measure the absolute magnitude and luminosity of galaxies. They all entail observational errors and limitations to a varying degree. The typical observational distance error $\Delta r$ is in the range $5 - 20 \%$, limiting the distance out to which useful peculiar velocity datapoints can be obtained.

The Tully-Fisher relation \citep{Tully1977} is an empirical relationship between the luminosity of a spiral galaxy and the amplitude of its gas rotation speed. This well-established method can provide distances with decent accuracy and a high data density over an appropriately large volume. The relation does not apply to elliptical galaxies, since they are in general not rotationally supported and contain few gas. In this case one can use the fundamental plane \citep{Faber1976,Djorgovski1987,Colless2001}, which establishes a relationship between the luminosity, the central stellar velocity, and the effective radius of the galaxy. An alternative is the surface brightness fluctuation (SBF) method \citep{Tonry2001}. However, he disadvantage of elliptical galaxies is that they are preferentially located in high-density regions (morphology-density relation, e.g.\ \citealt{Wel2010}) which do not sample the large scale galaxy flows. Other galaxy distance measurement methods include the Cepheid period-luminosity relation \citep{Freedman1990,Freedman2001}, and the tip of the red giant branch (TRGB) method \citep{Karachentsev2004,Rizzi2007}, although these methods suffer from limited reaches out to about only 10 -- 15 Mpc. Data out to very far distances and independent from the galaxy types can be obtained from observations of type Ia supernovae serving as standard candles \citep{Jha2007}. While this method is fairly accurate, it rests on serendipity and thus can provide only very sparse data samples. The Tully-Fisher method of measuring distances to spiral galaxies is the only one that currently combines all necessary assets: probing the space regions where coherent cosmic flows prevail and obtaining an adequate sampling of the Local Universe volume that would be suitable for a reconstruction of the underlying field and eventually the cosmological initial conditions.

If one wants to use this data as constraints for the cosmological ICs, it is important to understand the effect of the different observational errors, such as the knowledge of only one of the three velocity components, the limited distance out to which the data sampling extends, the distance errors, sparseness, and the sampling bias towards high-density regions. The current dataset of peculiar velocities used by the CLUES project is the Cosmicflows-1 catalogue. This data was assembled by \citet{Tully2008} and currently contains distances to 1797 galaxies in 742 groups, composed of different subsets obtained with different distance measurement methods and providing a fairly complete sampling of the sky within 3000 km/s. The ongoing observational work in the Cosmicflows program is currently directed towards preparing a much deeper and larger sample of peculiar velocities \citep{Courtois2011a,Courtois2011b,Courtois2012arXiv,Tully2012arXiv}. The upcoming Cosmicflows-2 catalogue will contain $\approx 7000$ distance measurements out to 6000 km/s with distance errors as low as 2\%, and eventually reach out to $15\,000$ km/s in the near future, exceeding all presently available data in both data volume and precision. It is of high importance to understand how to optimally use this improved data for constrained simulations, in what ways the quality of the simulations may be enhanced with the new data, and how to improve the constrained simulations method itself to optimally utilize the additional information provided by the datasets. 

This work is the second part of a series of papers devoted to this problem. In the first paper (Doumler et al. 2012, from here on Paper I) 
we presented the Reverse Zeldovich Approximation (RZA) method, which we use to generate constrained initial conditions from peculiar velocity data. As we showed in Paper I, RZA is a significant improvement over the original methodology developed by \citet{Hoffman1991}, \citet{Zaroubi1995,Zaroubi1999} and \citet{Klypin2003}. We refer the reader to Paper I for a thorough introduction to our method. Here, we investigate the effect of observational errors on the RZA reconstruction quality. For this, we use mock data drawn from a cosmological simulation.

This paper is organized as follows. In Section \ref{sec:mocks}, we briefly review the method we use to reconstruct the initial conditions from the data and present the set of mock catalogues that we use for our tests. In Section \ref{sec:analysis}, we perform the RZA reconstruction of ICs on the mock catalogues and analyze the reconstruction quality depending on different observational errors and biases. In Section \ref{sec:summary}, we summarize and discuss the obtained results.

%%%%%%%%%%%%%%%%%%%%%%%%%%%%%%%%%%%%%%%%%%%%%%%%%%%%%%%%%%%%%%%%%%%%%%%%%%%
\section{Method and test data}
\label{sec:mocks}

\subsection{RZA reconstruction}

The RZA is a Lagrangian reconstruction method, which applies the Zeldovich approximation backwards in time to peculiar velocity datapoints. It provides an estimator for the cosmic displacement field and the initial position of the observed object's progenitor in the linear regime at some early redshift $z_\tm{init}$ where we want to construct constrained initial conditions. The method is described in detail in Paper I; we give only a brief summary here.

Given a set of peculiar velocities $v_r^\tm{pec}$ with observational errors $\Delta v_r^\tm{pec}$, we first apply the Wiener Filter (WF) to the data \citep{Zaroubi1999}. With the WF, we can filter out the noise from observational errors and obtain an estimate $\bl v^\tm{WF}$ of the full three-dimensional velocity vector $\bl v$. Then, we estimate the cosmic displacement field $\bl \psi$ by extrapolating the linear-theory equation $\bl v = \dot a f \bl \psi$ to the current time of observation, $z=0$, and obtain the RZA estimate
\begin{align}
\label{eq:psi_rza}
\bl \psi^{\tm{RZA}} = \frac{\bl v^\tm{WF}}{H_0 f}\;\;\;.
\end{align}
We then continue with the Zeldovich approximation (ZA, \citealt{Zeldovich1970,Shandarin1989}), which relates the Lagrangian position $\bl q$ of a datapoint with the Eulerian position $\bl x(z)$ at redshift $z$ with the first-order\footnote{As we saw in Paper I, the first-order LPT is already capable of generating a very good estimate of $\bl \psi$ from $\bl v$, while being completely local and thus applicable to a sparse set of discrete data points. Higher-order LPT would unavoidably break this locality and require an integral of the full field over the whole volume of interest.}
Lagrangian perturbation theory (LPT) equation $\bl x(z) = \bl q(\bl x) + \bl \psi(z)$. Since we choose $z_\tm{init}$ such that the perturbations are very small, we approximate the initial position $\bl x_\tm{init}$ of the datapoint at $z_\tm{init}$ with $\bl q$. We then extrapolate the ZA to $z=0$ and apply it in time-reversed direction to obtain an estimate of $\bl x_\tm{init}$ for each datapoint,
\begin{align}
\label{eq:xinit_rza}
\bl x_{\tm{init}}^{\tm{RZA}}  = \bl r - \bl \psi^{\tm{RZA}}\;\;\;,
\end{align} 
where $\bl r$ is the observed position at $z=0$. This position can then be used to place constraints on the initial conditions and to run constrained simulations; the latter procedure will be analyzed in an upcoming paper, which will be Paper III in the series. In this work, we concentrate on the question how well we can recover $\bl x_\tm{init}$  for our data. A useful quantity in this context is the RZA error $d^\tm{RZA}$, which is the distance between the estimated and the actual initial position,
\begin{align}
\label{eq:rza_error}
d^{\tm{RZA}} = \left|\bl x_{\tm{init}} - \bl x_{\tm{init}}^{\tm{RZA}}\right| \;\;\;.
\end{align}
In our test setup, we can easily compute $d^\tm{RZA}$, since the mock data is drawn from a cosmological simulation of which the initial conditions are known. See Paper I for details on how this is accomplished in practice.

\subsection{Building the mock catalogues}

The procedure of generating realistic mock galaxy radial peculiar velocity catalogues was described in detail in Paper I. Here, we give only a brief summary.

We use the BOX160 simulation \citep{Cuesta2011}, a constrained simulation of the Local Universe in a volume of boxsize 160 Mpc/$h$, as the test universe. We build a dark matter halo catalogue at $z=0$ with the halo finder AHF \citep{Knollmann2009}, assume that each galaxy sits at the centre of a dark matter halo, and take the peculiar velocity of the halo as a proxy for the galaxy peculiar velocity. We consider only its radial component relative to an arbitrarily chosen mock observer, and add observational errors by assuming Gaussian-distributed relative distance errors $\delta r$ with some fixed standard deviation $(\delta r)_\tm{rms}$. We further have to compensate for the fact that Lagrangian perturbation theory breaks down on small scales where shell crossing occurs. In particular, it does not account for the dynamics within virial haloes. To address that limitation, we are using here only `parent' haloes as tracers of the large-scale velocity field and ignore substructures. `Parent' haloes are defined here as virial haloes that are not contained, fully or partially, within more massive haloes.

\subsection{The mock catalogue set}

\begin{table*}
\begin{tabular}{ccccccccl}
\hline \hline
mock & constraint & $R_{\tm{max}}$ & $\delta r$ & mass cut & halo selection & $M$ & $\sigma_{\tm{NL}}$ \\ 
name & type & [Mpc/$h$] & & $[\log M/M_\odot]$ & & & [km/s]  \\[2mm]
 \hline 
C30\_00 & $v_r$ & 30 & 0 \% & $>11.9$ & by mass & 588 & 221 & \rdelim\}{5}{0mm}[ I]  \\ 
C30\_05 & $v_r$ & 30 & 5 \% & $>11.9$ & by mass & 588 & 228 \\ 
C30\_10 & $v_r$ & 30 & 10 \% & $>11.9$ & by mass & 588 & 235 \\ 
C30\_15 & $v_r$ & 30 & 15 \% & $>11.9$ & by mass & 588 & 242 \\
C30\_20 & $v_r$ & 30 & 20 \% & $>11.9$ & by mass & 588 & 246 \\[2mm]  

A30\_10 & $v_r$ & 30 & 10 \% & $>12.3$ & by mass & 282 & 242 & \rdelim\}{4}{0mm}[ II]  \\  
B30\_10 & $v_r$ & 30 & 10 \% & $>12.1$ & by mass & 413 & 235 \\ 
D30\_10 & $v_r$ & 30 & 10 \% & $>11.7$ & by mass & 898 & 216 \\
E30\_10 & $v_r$ & 30 & 10 \% & $>11.5$ & by mass & 1243 & 198 \\[2mm] 

C40\_10 & $v_r$ & 40 & 10 \% & $>11.9$ & by mass & 1256 & 179 & \rdelim\}{3}{0mm}[ III]  \\ 
C50\_10 & $v_r$ & 50 & 10 \% & $>11.9$ & by mass & 2184 & 176 \\ 
C60\_10 & $v_r$ & 60 & 10 \% & $>11.9$ & by mass & 3518 & 190 \\[2mm]
  
E40\_10 & $v_r$ & 40 & 10 \% & $>11.5$ & by mass & 2614 & 156 & \rdelim\}{3}{0mm}[ IV]  \\  
E50\_10 & $v_r$ & 50 & 10 \% & $>11.5$ & by mass & 4701 & 154 \\
E60\_10 & $v_r$ & 60 & 10 \% & $>11.5$ & by mass & 7637 & 165 \\[2mm]

L30\_10 & $v_r$ & 30 & 10 \% & $<11.9$ & by mass & 588 & 208 & \rdelim\}{3}{0mm}[ V]  \\   
I30\_10 & $v_r$ & 30 & 10 \% & --- & by isolation & 588 & 180 \\ 
R30\_10 & $v_r$ & 30 & 10 \% & --- & random & 588 & 219 \\[2mm]

C30\_3D & $v_x, v_y, v_z$ & 30 & 0 \% & $>11.9$ & by mass & 1764 & 183  & {-- } VI \\ [2mm]
\hline \hline
\end{tabular}
\caption{Overview over the mock catalogues extracted from the BOX160 simulation that were used for the reconstructions. From left to right: an abbreviation used in this chapter to refer to each mock and the reconstruction computed from it; the physical quantity constrained by each data point (either only the radial component $v_r$ or all three cartesian components of the halo peculiar velocity); the radius of the spherical data zone in Mpc/$h$; the mock rms distance error that was added; the mass limit above which haloes are selected; the selection method; the total number $M$ of constraints (for $v_r$ this is equal to the number of data points); and the $\sigma_{\tm{NL}}$ parameter that was used for each Wiener Filter reconstruction to enforce $\chi^2/\tm{dof} = 1$.}\label{table:vz0recon}
\label{table:mocks}
\end{table*}

The set of mock catalogues presented here is an expansion of the data that we introduced in Paper I. From the BOX160 AHF halo catalogue, we created a total of 20 different mock peculiar velocity catalogues in order to test how the observational distance error, the amount and distribution of data points, and the size of the observational volume are going to affect RZA reconstruction. Table \ref{table:mocks} summarises the basic parameters of all mocks used in this paper. Each of the mocks is referred to by a name encoding its properties. The first letter characterises the method of halo selection (A -- E: by mass cut; L,I,R by other criteria); the next two digits show the radius of the observational volume $R_{\tm{max}}$ in Mpc/$h$; and the last two digits are the rms distance error $(\delta r)_\tm{rms}$ in percent. The last two mock catalogues do not feature distance errors but instead contain 3D peculiar velocity data, which will be discussed below. In this case the last two digits are 3D.

The ''fiducial'' catalogue is the C30\_10, which we consider a ``typical'' sparse peculiar velocity dataset. We take the procedure of considering only main haloes as a proxy for the ``grouping'' performed on observational data. The C30\_10 contains all main haloes above a mass cut $M_\tm{min} = 10^{11.9} M_\odot/h$ within $R_{\tm{max}} = 30$ Mpc/$h$, yielding 588 radial velocity datapoints. This choice gives the C30\_10 similar properties to the grouped Cosmicflows-1 catalogue (see Paper I for details). The fiducial choice of 10\% rms distance error is also similar to the observational data: while the median rms distance error is somewhat higher at 13\% on the individual galaxies in Cosmicflows-1, this error reduces when the galaxies are arranged in groups. The C30\_10 is interesting because even if the data are improving in terms of the number of individual galaxy distances, the number of galaxy groups in a radius of 30 Mpc/$h$ is probably not going to vary by much, nor is the accuracy on the most nearby distances.

With the C30\_10 mock as the starting point, we vary the rms distance error in five steps between none and 20\%, yielding the mocks C30\_00 through C30\_20. We also vary the mass cut from $M_\tm{min} = 10^{12.3} M_\odot/h$ to the minimum of $10^{11.5} M_\odot/h$ in five steps, yielding the mocks A30\_10 through E30\_10. A30\_10, the sparsest sample, has the fewest data points of all mocks with only 282 radial velocities. 
We also construct mocks with larger observational volumes around $\bl r_{\tm{MW}}$, varying $R_{\tm{max}}$ from 30 to 60 Mpc/$h$ in four steps, for two different mass cuts, yielding the mocks C30\_10 through C60\_10 and E30\_10 through E60\_10. Although the RZA method itself places no restrictions on the allowed size of the data volume, we do not consider $R_{\tm{max}}>60$ Mpc/$h$ here in order to avoid problems with the periodic boundary conditions of the box. The E60\_10 mock has the most data points with 7637 radial velocities. We estimate that this mock is comparable to the upcoming Cosmicflows-2 catalogue in terms of data quality. 
 
Next, we want to explore how other halo selection criteria rather than a mass cut will affect the reconstruction. 
Regarding again the C30\_10 mock as a ``fiducial" one, we fix the amount of data points constant at 588, as well as the distance error and the data volume. Then, for the L30\_10 mock, we take the 588 next most massive points after the ones in C30\_10, so that they all have a mass below $10^{11.9} M_\odot/h$, to check the reconstruction quality if only less massive objects are considered.  For the I30\_10 mock, we consider the 588 most isolated objects leading to a yet different sampling of the same volume. We define the isolation radius $R_{iso}$ of a halo as the distance to the next more massive halo, or in other words, the radius within which a halo is the most massive. We then choose the 588 objects with the largest $R_{iso}$ within 30 Mpc/$h$, which corresponds to all haloes within 30 Mpc/h with $R_{iso} >  2.1$ Mpc/$h$.
In the last variation of data selection, we randomly pick 588 points from all main haloes in the data volume, regardless of their mass or other properties, yielding the R30\_10 mock. Randomly picking haloes mimics the observational data of spiral galaxy peculiar velocities obtained with the Tully-Fisher method, which are not located at the highest density peaks of the galaxy distribution (as the haloes selected by mass), but are selected on the random basis of their inclination on the sky being grater than 45 degrees.

Finally, we want to quantify by how much the reconstruction quality is degraded by the fact that only the radial component $v_r$ is observable, rather than the full three-dimensional velocity vector $\bl v$. This is interesting, since in the future the transverse peculiar velocities of galaxies may become accessible to observations as well \citep{Nusser2012arXiv}. We construct the mock C30\_3D, which has its data points at the same positions as the C30\_00 mock, but for each object it lists all three components $v_x, v_y, v_z$ of the velocity instead of $v_r$.

Of course, there are more observational features that could be incorporated in the mocks. For example, one could add a Zone of Avoidance (ZoA). However, it has been already established that the Wiener Filter mean field successfully extrapolates into such unsampled regions and handles datasets well that are sparse in an inhomogeneous and/or anisotropic way \citep{Courtois2012}. We already exploit this behaviour by using sparse mock catalogues with a very limited observational volume and huge gaps in the data in underdense regions; adding an additional gap will not fundamentally change the situation. We confirmed this by performing tests with mocks featuring a ZoA of $\pm 11^\circ$, which produced practically the same result. Additionally, it would be possible to incorporate  the observational bias due apparent magnitude selection effect into the mock data, or modelling the different galaxy types and measurement methods in more detail. For this, it would be necessary to populate the haloes with galaxies of different morphologies and luminosities by setting up a full semianalytic model on top of the $N$-body simulation. This would make the situation unnecessary complex without additional insight into the validity of the RZA and our method of generating constrained initial conditions. We therefore defer such in-detail studies to future work.

%%%%%%%%%%%%%%%%%%%%%%%%%%%%%%%%%%%%%%%%%%%%%%%%%%%%%%%%%%%%%%%%%%%%%%%%%%%
\section{Analysis and results}
\label{sec:analysis}

\subsection{Reconstruction quality}

\begin{figure}
\centering
\includegraphics[scale=0.7]{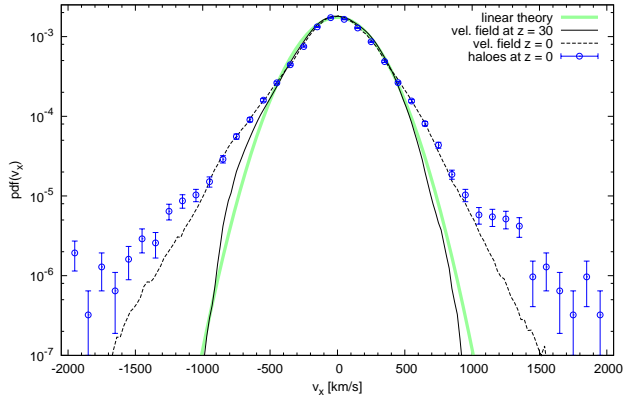}
\caption{Distribution function of one velocity field component (vx) in BOX160 at the initial conditions
z = 30 and from the simulation snapshot at z = 0 (normalised to the same growth rate). The
blue points show the distribution function of dark matter halo velocities at z = 0. The green curve shows
a theoretical Gaussian distribution, with the mean and standard deviation adjusted to $<v_x>$ and $<|v_x|>$,
respectively.}
\label{fig:histogram_v}
\end{figure}

%%%%%%%%% added in revised version
The method we use rests on the assumption that the peculiar velocity field of the Local Universe, observed through peculiar velocities of galaxies, does not deviate too much from the linear velocity field of the ICs at some early redshift $z_\tm{init}$, so that we can directly use the values of the radial peculiar velocities $v_r^{\tm{pec}}$
as constraints $c_i$. This approach differs from the WF reconstruction from such data by the requirement that the result must be a linear Gaussian random field suitable for initial conditions. This is possible, because the peculiar velocities of dark matter haloes at z=0 do not deviate too much from the linear Gaussian distribiution at initial redshift $z_\tm{init}$ (see Figure~\ref{fig:histogram_v}). One obvious source of incompatible non-linearities in the data are the virial motions of galaxies gravitationally bound to larger objects such as galaxy clusters. This can be overcome by an appropriate grouping of the data points, which effectively ÒlinearisesÓ the data. Virial motions are, of course, not the only source of non-linearity. Another example of non-linearities in the observable peculiar velocity field at $z=0$ is the general enhancement of halo peculiar velocities due to local overdensities. Additionally, with the haloes we are sampling the peculiar velocity field at the positions of the density peaks, which adds another bias (see \cite{Bardeen1986}). The combination of these effects leads to the non-linear tails of enhanced halo velocities (blue points) at the high-velocity ends visible in Figure 1. See \cite{Sheth2001b} and \cite{Hamana2003} for a more detailed discussion of these effects.\\

However, the non-linearities can be compensated for in a statistical sense. Considering the non-linear effects as a form of statistical scatter, one can add a new non-linearity term  to the WF autocorrelation matrix of the data:   $\sigma_{\tm{NL}}$ \citep{Bistolas1998}:
As explained in Paper 1 (Doumler et al. 2012) of this serie, the value of $\sigma_{\tm{NL}}$ is chosen when $\tilde{\chi}^2=1.0$. For observational radial peculiar velocity data, we find a typical value of $\sigma_{\tm{NL}} \approx 200$ km/s. The initial conditions are then created as follows. First one constructs and inverts the autocorrelation matrix of the data. Then an appropriate boxsize must be chosen such that the data zone lies well within the computational volume. Then, the WF/CR operator can be evaluated on this volume, leading to a linear density field $\delta_0^{\tm{CR}} (\bl x)$. This field can then be scaled with the growth factor $D_+$ to the desired starting redshift $z_\tm{init}$ of the simulation (\ref{eq:ansatzforlineardensity}) and used to set up $N$-body ICs.
\begin{align}
\label{eq:ansatzforlineardensity}
\delta(\bl x, t) = D (t) \; \delta_0 (\bl x) \;\;\; ,
\end{align}
In the linear approximation the initial shape $\delta_0 \bl( x)$ of the overdensity distribution remains fixed, and its amplitude scales with time proportional to the factor $D(t)$\\

%%%%%%%%% added in revised version

\begin{figure}
\centering
\includegraphics[scale=1.1]{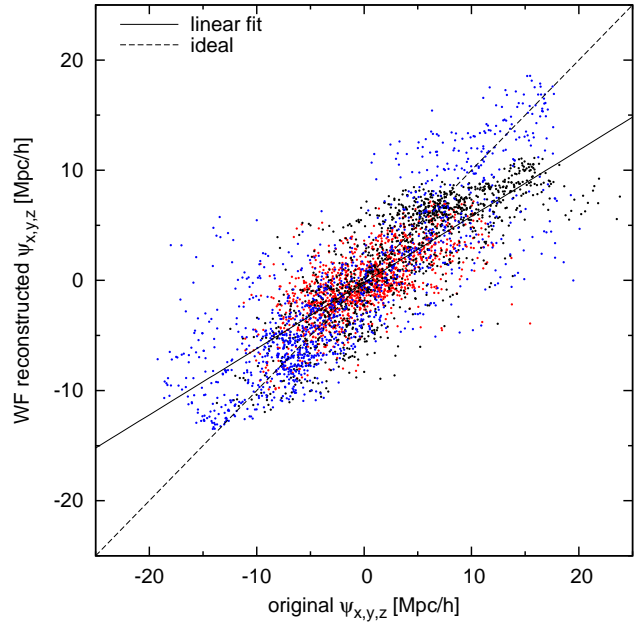}
\caption{Comparison of WF reconstructed vs. actual displacement $\bl \psi$ for all haloes within 30 Mpc/$h$ of the observer at the discrete positions of those haloes, using the C30\_10 mock for the reconstruction. The solid line shows a linear regression fit $\psi_i^{\tm{WF}} = \beta \cdot \psi_i^{orig} + \varepsilon$; the dashed line would be the ideal result $\psi_i^{\tm{WF}} = \psi_i^{orig}$; where $i \in \{ x,y,z\}$ are the three cartesian components $x$ (red), $y$ (black), $z$ (blue).}
\label{fig:psicheck_paper2}
\end{figure}

After reconstruction of the three-dimensional peculiar velocity values $\bl v^\tm{WF}$ with the Wiener Filter, we obtain the reconstructed cosmological displacements $ \bl \psi^{\tm{WF}}$ of the datapoints with Eq. \ref{eq:psi_rza}. We compare the displacements $\bl \psi$ of all main haloes within the $R_{\tm{max}}=30$ Mpc/h mock volume with the values of the reconstructed displacement field $ \bl \psi^{\tm{WF}} $ at the positions of those haloes. Such a scatter plot is given in Figure \ref{fig:psicheck_paper2} for the C30\_10 reconstruction. Here, the average rms error per component is 4.17 Mpc/$h$, the slope is 0.64, and the correlation factor is 0.82. The correlation is significantly poorer than the one between $\bl v^\tm{WF}$ and the true $\bl v$, since the RZA error (the intrinsic scatter between $\bl \psi$ and $\bl v$ due to deviations from first-order LPT) is added on top of the error from the imperfect WF reconstruction. Still, we obtain a reasonable correlation.

\begin{figure*}
\centering
\includegraphics[scale=0.95]{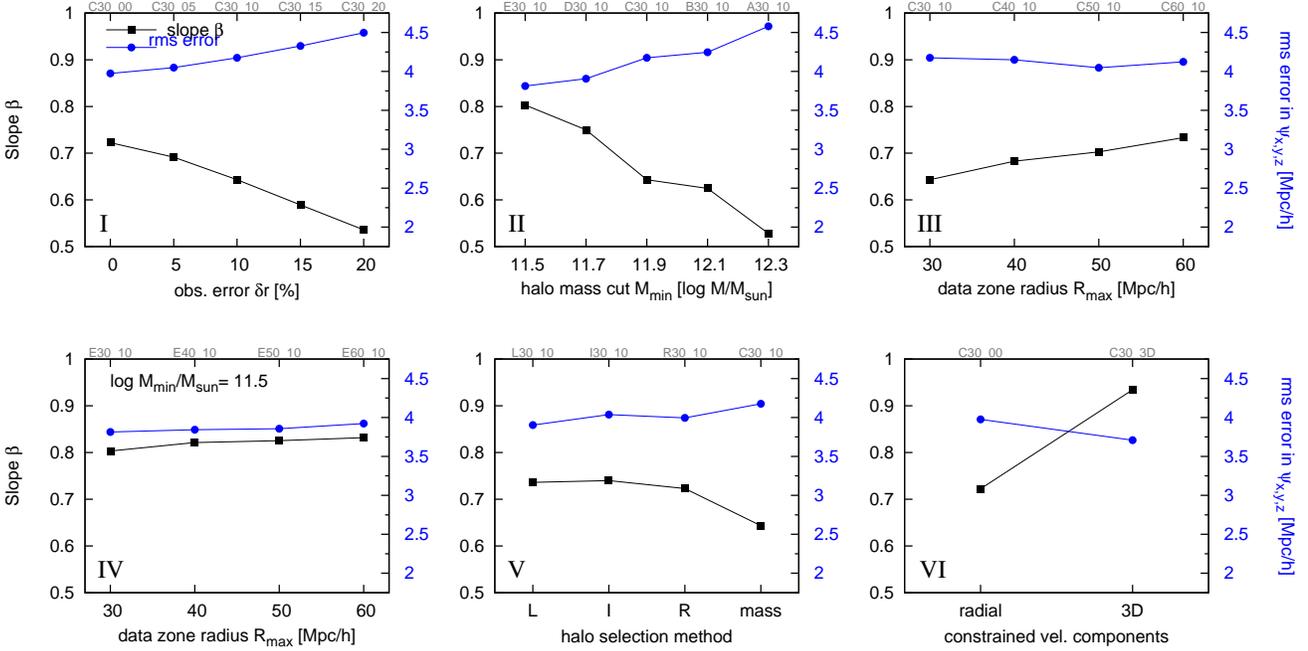}
\caption{Slope $\beta$ (black; left scale) and rms error (blue; right scale in Mpc/$h$) for a linear regression comparison of the reconstructed displacement field $\bl \psi^\tm{RZA}$ with the original $\bl \psi$ within 30 Mpc/$h$ of the observer.
}
\label{fig:linreg}
\end{figure*}

We now perform the same procedure with all 20 mocks. We perform the Wiener Filter reconstuctions with our \icecore\ code (see Paper I), after determining the appropriate ``non-linearity regularization'' parameter $\sigma_{\tm{NL}}$ (cf. Table \ref{table:mocks}). Considering only the result inside the 30 Mpc/$h$ sphere, we then compare the resulting $ \bl \psi^{\tm{WF}}$ values   with those of the true $\bl \psi$ from the original simulation. This is fitted with a linear regression line as in Figure \ref{fig:psicheck_paper2}. However it would be tedious to handle such a huge amount of scatter plots. To present the results in a more compact way, we concatenate all three cartesian components for the linear regression and consider the resulting slope and rms error per component averaged over all three components. Figure \ref{fig:linreg} shows the slope (black) and rms error (blue) for the reconstructed halo displacements $\bl \psi^{\tm{WF}}$ vs. the true $\bl \psi$. 

The different panels in Figure \ref{fig:linreg} divide the mock reconstructions into groups by the different mock observational parameters that are varied. The roman numerals correspond to the mock groups in Table \ref{table:mocks}. Panel I shows the dependance of the reconstruction quality on the rms distance error; panel II varies the mass cut (and therefore the amount and density of data points) for the halo mass-selected mocks; panel III shows the effect of increasing the data volume beyond the default $R_{\tm{max}}=30$ Mpc/$h$ for the default mass cut at 11.9; panel IV repeats the same for lower mass cut mocks at 11.5; panel V shows mocks with different selection methods while keeping the number of datapoints constant; finally, panel VI compares radial with three-dimensional input data.

%\begin{figure*}
%\centering
%\includegraphics[scale=1.1]{figures/smoothcheck.eps}
%\caption{Cell-by-cell comparison of WF reconstructed vs. actual velocity field $\bl v(\bl r)$ within 30 Mpc/$h$ of the observer, using the C30\_10 mock for the reconstruction. Top row: Without smoothing. Bottom row: with 5 Mpc/$h$ Gaussian smoothing on both fields. The solid line shows a linear regression fit $v_i^{\tm{WF}} = \beta \cdot v_i^{orig} + \varepsilon$; the dashed line would be the ideal result $v_i^{\tm{WF}} = v_i^{orig}$.}
%\label{fig:smoothcheck}
%\end{figure*}

%\subsection{Reverse Zeldovich Approximation}

%Figure \ref{fig:linreg}, Figure \ref{fig:kernelpdf}, Figure \ref{fig:binnedmodels}, Figure \ref{fig:viewingangle}. %, Figure \ref{fig:mockrzamap}.

In general, the slope $\beta$ is a measure for how well the underlying field is constrained by the data. Ideally it should be 1; the WF introduces a filtering bias due to its conservative nature, and reduces $\beta$ to a value $< 1$. The rms error is a measure for how well the WF result reproduces the ``true'' solution for the halo displacements.

We have to distinguish two different scenarios. Generally, if $\bl v^\tm{WF}$ is compared to the true $\bl v$ at $z=0$, the reconstruction can become almost arbitrarily accurate if we sufficiently increase the data quality (it is well established that the WF is a very good filter in this case; see \citealt{Zaroubi1999}). For RZA however, we want to estimate the halo displacements in order to apply the RZA reconstruction. The disagreement between $\bl \psi^\tm{WF}$ and $\bl \psi$ however is dominated by the RZA error rather than by the WF reconstruction quality; varying the mock properties has a lesser effect. There is a ``wall'' at around 3.5 Mpc/$h$ rms error per component that cannot be penetrated even with the best-quality mocks. This is the scale at which, averaged over the mock volume, the $\bl \psi$ and $\bl v$ fields themselves disagree, because the quasi-linear assumption is not valid on these scales. Of course, this disagreement varies from region to region and highly depends on the local density (amongst other quantities), as shown in Paper I.

\subsubsection{Distance errors and data sparseness}

As expected, the distance errors and the mass cut both have a significant influence on the quality of the reconstruction (groups I and II). The latter seems to be more important: starting from the C30\_10 mock, a higher improvement is obtained when the number of data points is increased than if the distance errors are decreased. This is interesting when considering the upcoming observational data, where the sparseness of the data will be reduced more effectively than the observational distance errors compared to present observations. The next lower mass cut mock D30\_10 with 10\% distance errors but 898 radial velocities instead of 588 gives a better reconstruction than the mock that keeps the 588 points but has no distance errors at all. This is remarkable because an rms error of 10\% in distance at a distance of 30 Mpc/$h$ leads to a rms error of 300 km/s on the radial velocities, and because of the Gaussian distance error distribution, some velocities in the mocks have error bars up to 100\% and higher. It demonstrates that due to their coherence on large scales, peculiar velocities are an excellent input data source for the WF despite the large errors: even a few coherent data points with 100\% errors and separated by a few Mpc/$h$ represent a strong measurement of the local velocity field.

Increasing the data volume (groups III and IV) has a much lesser effect on the reconstruction quality within the inner volume of 30 Mpc/$h$ radius. This is expected for peculiar velocity data as opposed to redshift data, where a larger data volume would lead to a significantly better overall result. If we increase the total volume of the mock catalogue out to a distance of 60 Mpc/$h$, the improvement on the reconstruction inside the 30 Mpc/$h$ is minimal. There is some effect for the high 11.9 mass cut mocks, since the additional information partly compensates for the sparse samling. But for the low mass cut mocks at 11.5 there is no significant improvement, although the E60\_10 mock contains already 7637 data points in total. This reflects a known favourable property of the WF: it successfully reconstructs the tidal component of the velocity and displacement fields, i.e.\ the part that is induced by the mass distribution outside the data zone (e.g.\ \citealt{Courtois2012}). This also includes the dipole term, i.e.\ the bulk motion of the data volume due to the external field, which in our case is significant\footnote{This external field corresponds to the large-scale Local Flow in that is observable in the Local Universe, see \citet{Courtois2012}.}. Adding more detail on this outside field does not significantly change the tidal component on the inner volume. The RZA reconstruction with WF can therefore work with small data volumes compared to density-based methods (i.e.\ based on redshift data). Any such method is by design not able to reconstruct the tidal component: information outside of the data zone cannot be inferred from galaxy positions alone. A density-based Lagrangian reconstruction from a catalogue volume of only 30 Mpc/$h$ would therefore be of little use. On the other hand, Lagrangian reconstruction from peculiar velocities are an ideal tool to study the tidal flows on large scales.

\subsubsection{Selection criteria and connection to distance measurement methods}

Panel V in Figure \ref{fig:linreg} compares mocks created from different data selection criteria while keeping the number of data points constant at 588 inside 30 Mpc/$h$. The mass cut C, which picks the 588 most massive objects, is compared against selecting lower-mass objects (L), the most isolated (I) and randomly picked (R) objects. It is very interesting to observe here that any of the alternative selection methods works better than selecting by mass. Mass selection is biased towards a higher sampling the overdense regions and a poorer sampling of the less dense regions. Any other selection will sample the less dense regions more completely and therefore conserve more information about the large-scale modes of the cosmic matter distribution. Additionally, a more homogeneous sampling not biased towards the denser regions will be less affected by the non-linear enhancement bias of the peculiar velocity field \citep{Sheth2001b,Hamana2003}. Therefore, such a sampling  creates a WF solution that is better constrained and has a lower rms error. This detail is interesting in the context of using different observational distance measurement methods to obtain input data. 

The interpretation is that a galaxy sample more evenly distributed around the sky and sufficiently probing less dense regions would lead to better reconstrutions than a galaxy sample preferentially located in dense regions. It suggests that galaxy distance samples of spiral galaxies (such as those derived from the Tully-Fisher relation) may be more ideal for RZA reconstruction than those primarily containing early types, which are biased towards massive surrounding haloes and dense environments. The randomly selected mocks (R) mimic the observational data of spiral galaxy peculiar velocities which are not located at the highest density peaks of the galaxy distribution, but are selected on the random basis of their inclination on the sky being grater than 45 degrees (this is an absolute requirement for the Tully-Fisher method). Similarly, the lowest-mass-selected mocks (L) mimic preferring less massive spiral galaxies over more massive elliptical galaxies, and the isolation-selected mocks (I) mimic preferring galaxies in less dense environments. All three selection criteria lead to a similar result and provide a better reconstruction than focussing on the most massive objects.

\subsubsection{Radial vs. 3D data}
\label{subsubsec:radialvs3d}

Group VI addresses the question how much of the information is missed due to the fact that only radial components of the velocity are observable. For this sake, we constructed the C30\_3D mock, pretending that the full 3D velocity vector would be accessible in some way. It has three-dimensional velocity data on the usual 588 haloes inside 30 Mpc/$h$. In C30\_3D, there are no added errors because it is not yet demonstrated with observational data, how one should derive mock 3D velocity errors from a distance error. 
 This will happen in the future through direct astrometric measurements. \citep{Nusser2012arXiv} gives an estimate of this error. Naively, it can be
expected that the errors scales linearly with distance: the velocity is proportional to the product
of the angular distance and the angle between two repeated observations of the same object. An
error on the angle yields an error on the velocity proportional to the distance. There is also a
component of the error introduced by an error on the distance: but this one is also typically
proportional to the distance (as indicated in this paper).
We therefore compare it to C30\_00, which has no errors either. Surprisingly, for the displacement reconstruction, the additional 3D information has much less effect than we would have expected, because the total scatter is dominated by the RZA error. The WF filtering bias ($\beta < 1$) is almost completely removed by adding three-dimensional information to the data, but the rms error on the displacement components is reduced only from 4.0 to 3.7 Mpc/$h$. This means that even for very high-quality data, the RZA estimate of the displacement $\bl \psi$ is still significantly limited in precision by the disagreement between the large-scale velocity field and $\bl \psi$ due to non-linear motions that do not follow the Zeldovich approximation.

\begin{figure*}
\centering
\includegraphics[scale=0.7]{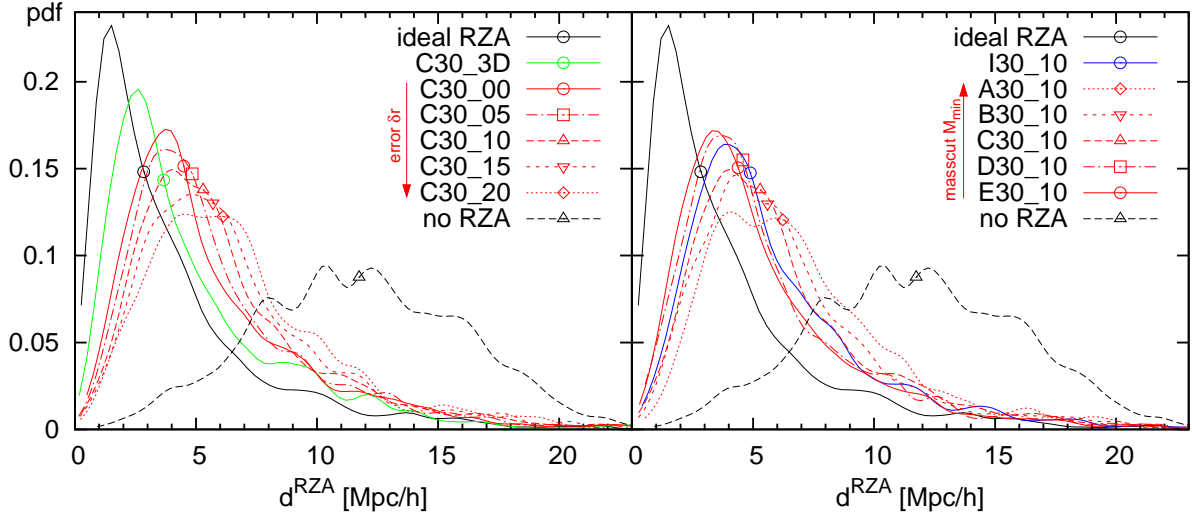}
\caption{Probability distribution function (pdf) of the RZA displacement error $d^{\tm{RZA}}$ for different mock reconstructions, as well as a theoretical reconstruction using the exact 3D halo velocity of all haloes within the data zone, i.e.\ $d = |\bl \psi - \bl v/H_0f|$ (black solid), and no reconstruction at all, i.e.\ $d = |\bl \psi|$ (black dashed). The symbols are placed at the median of each distribution. Each pdf was convolved with a 0.5 Mpc/$h$ Gaussian kernel to obtain a smoother plot.}
\label{fig:kernelpdf}
\end{figure*}

\begin{figure*}
\centering
\includegraphics[scale=1.08]{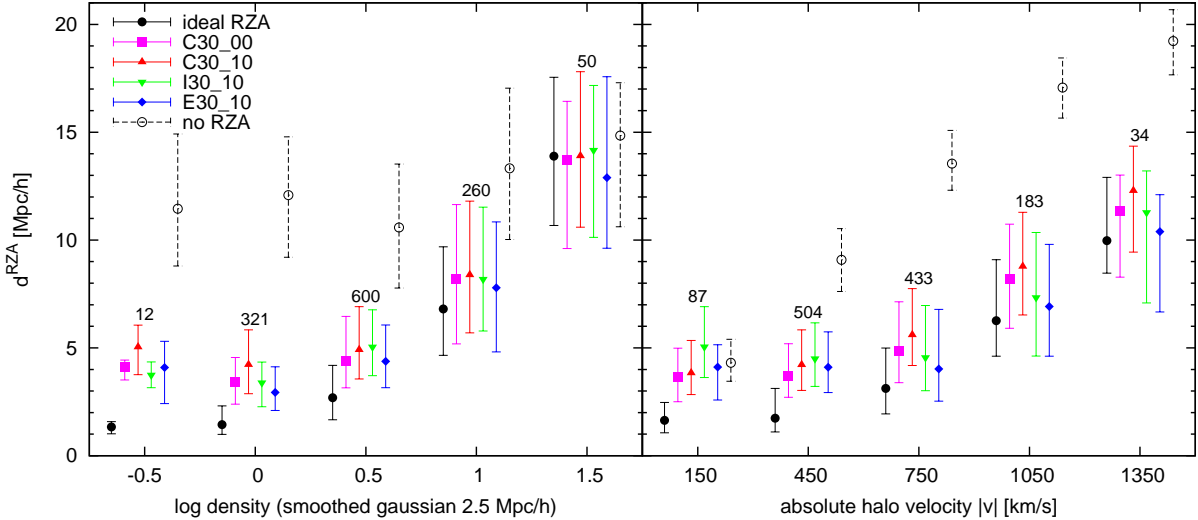}
\caption{RZA displacement error $d^{\tm{RZA}}$ of haloes inside the data zone binned over the underlying density and the halo velocity for different mock reconstructions, as well as the ideal RZA (solid black) and no RZA (dashed black). The number of haloes in each bin is also given.}
\label{fig:binnedmodels}
\end{figure*}

\begin{figure*}
\centering
\includegraphics[scale=1.08]{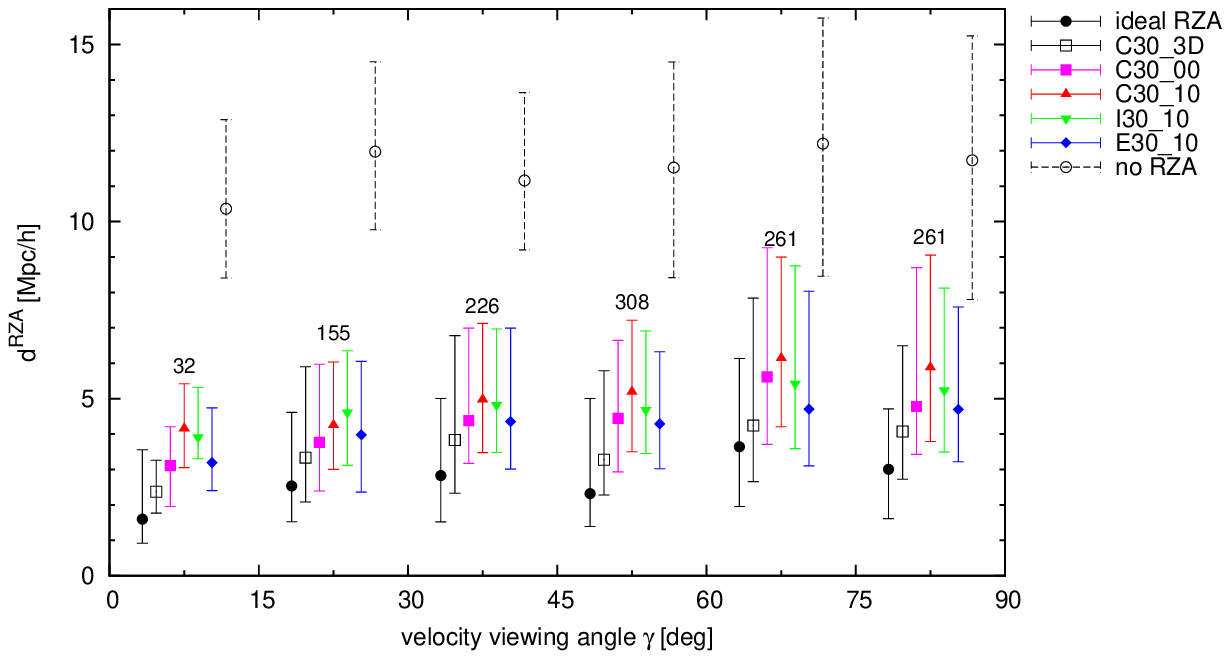}
\caption{RZA displacement error $d^{\tm{RZA}}$ of haloes inside the data zone binned over the peculiar velocity viewing angle $\gamma$, i.e.\ the angle between the full 3D velocity of the halo $\bl v$ and its observed radial component. The C30\_3D reconstruction utilises full three-dimensional data and thus does not depend on $\gamma$, as well as the ideal RZA and no RZA.}
\label{fig:viewingangle}
\end{figure*}

%\begin{figure*}
%\centering
%\includegraphics[scale=0.6]{figures/mockrzamap.eps}
%\caption{Actual BOX160 displacement field (left) vs. RZA reconstruction from radial peculiar velocities (using mock E30\_10). The dots show the $z=0$ position $\bl r$ of the haloes contained in the mock data; the lines show the displacement vector $\bl \psi$ from the initial conditions. In the reconstruction, the RZA error $d^{\tm{RZA}}$ is highlighted by colour-coding of the points. The contours show the underlying dark matter density (smoothed with a 1.5 Mpc/$h$ Gaussian). Shown is a 15 Mpc/$h$ thick slice containing the simulated Local Supercluster with Virgo at X=82, Y=74. The red cross marks the position of the mock observer.}
%\label{fig:mockrzamap}
%\end{figure*}

\subsection{RZA error distribution}
\label{subsec:rzarecon_rzaerrorpdf}

\subsubsection{General}

To obtain a more detailed view at the reconstruction quality, we consider the displacement error $d^{\tm{RZA}}$ for the haloes inside the mock volume. Figure \ref{fig:kernelpdf} shows the distribution of $d^{\tm{RZA}}$ for different mocks. Recall that $d^{\tm{RZA}}$ is computed for \emph{all} haloes inside the mock volume, regardless of whether they were part of the particular mock from which a reconstruction was computed, so that we can directly compare the overall reconstruction quality from all mocks.

Compared to the distribution of halo position error without RZA reconstruction (black dashed line), the RZA gives a significant improvement. While half of the haloes have an $x_{\tm{init}}$ position error above 11 Mpc/$h$ without RZA, this is the case for only around 10\% of the haloes with RZA reconstruction. Used as constraints for initial conditions, this leads to much more exact constrained simulations (we refer the reader to the upcoming Paper III for more details). The distribution of course depends on the quality of the mocks. Typically, the reconstructions from radial velocities have median $d^{\tm{RZA}}$ values around 5 Mpc/$h$ and a skewed $d^{\tm{RZA}}$ distribution. For comparison we plot also the ``ideal'' RZA (solid black), directly using the exact known 3D halo velocities $\bl v$ at $z=0$ to estimate $\bl \psi$ as $\bl v / H_0 f$. As already mentioned in Paper I, this gives a median $d^{\tm{RZA}}$ of 2.8 Mpc/$h$ inside the mock volume and a similarly skewed $d^{\tm{RZA}}$ distribution. The mock groups I and II are represented in red lines in the left and right panels of Figure \ref{fig:kernelpdf}, respectively.

The trend of degrading quality with increasing distance errors and decreasing number of datapoints is repeated here. However, the differences between the different $d^{\tm{RZA}}$ distributions do not seem very huge considering that the respective quality of the mocks  differ from each other considerably: the distance errors are varied between none and 20\%, and the amount of datapoints inside the same volume between 282 and 1243. Comparing the mass cuts in the right panel, there is a hint that the reconstruction quality starts to saturate: the difference between the D30\_10 and E30\_10 catalogues is relatively small. Indeed, if the data quality increases, the overall reconstruction error is more and more dominated by the RZA error. The RZA reconstruction quality would thus not significantly increase if we add data mapping scales below the $d^\tm{RZA}$ scale. This critical scale, below which the RZA reconstruction breaks down, depends on the local density and is 3.5 Mpc/$h$ per cartesian component on average. Therefore, about one constraint per $(3.5\;\tm{Mpc}/h)^3$ on average would provide a sufficient data density.

The blue line in the right panel uses the I30\_10 mock, with 588 points like the C30\_10 mock (red long dashed) but more evenly distributed, which shows a noticeable improvement in the $d^{\tm{RZA}}$ distribution. The green line in the left panel uses the three-dimensional data of C30\_3D. The distribution is much closer to the ``ideal RZA'' than all the radial velocity mocks. We can derive from this that for RZA reconstruction, actually more accuracy is lost by having only the radial component than suggested by comparing the rms errors on $\bl \psi$ (section \ref{subsubsec:radialvs3d}).

Figure \ref{fig:binnedmodels} is conceptually similar to Figure 4 in Paper I,
% REMEMBER TO CHANGE THE FIGURE NUMBER IN CASE PAPER I CHANGES!!!
but comparing different reconstructions against each other. The $d^{\tm{RZA}}$ error is binned against the underlying density and the absolute velocity. The fiducial C30\_10 mock (red) is contrasted with the I30\_00 mock having a more uniform data distribution (green), the E30\_10 mock having about twice the amount of data points (blue), and the C30\_00 mock having no data errors (purple). ``Ideal'' RZA (black) and no RZA (dashed black) is included for comparison. One can again see the main result: no matter how the data quality is in detail, the RZA reconstruction greatly reduces the distance errors for most of the objects in the data compared with no Lagrangian reconstruction of their initial positions (dashed black). The left panel of Figure \ref{fig:binnedmodels} highlights again the strong dependance of $d^{\tm{RZA}}$ on the underlying density. For the displacement field in underdense regions, it seems particularly important to have a more homogeneous sampling of the data, mapping this region well and with a distance error as low as possible: the C30\_10 mock performs noticeably worse here than both the more homogeneous isolated halo mock I30\_10 and the mock C30\_00 with no errors. Conversely, in high-density regions, $d^{\tm{RZA}}$ is high and errors on the individual velocities have little influence, since those velocities do not track the displacement field well due to the high non-linearity. In this case, it seems better to have more data points in general: the E30\_10 mock performs somewhat better than the others. Fed with more data, the WF algorithm has more chance to filter out the noise. It is interesing that in the densest bin (around 1.5 smoothed density), the WF result from E30\_10 performs even better than the theroretical ``ideal'' RZA. This can be understood from the properties of WF as a linear filter. The true halo velocities, even when perfectly known, are a bad tracer of the displacement field in the densest regions. But if the surrounding field is mapped sufficiently well, the WF can create a solution for the displacement field that is closer to the more linear actual displacements. The difference is very small though. We can argue that when there is a subsequent step of Wiener filtering on the velocity data, rigorously identifying and removing substructure is less critical than it appeared in Paper I. More important is the insight that, since reconstruction quality from the different mocks does not differ as much as one could expect, the procedure of Wiener filtering and subsequent RZA reconstruction is restricted more by the limitation of the scheme itself because of the underlying non-linearity of the system, and to a lesser degree by the actual data quality.

\subsubsection{Viewing angle}
\label{subsec:rzarecon_3d}

It is worth to analyse in more detail the impact of having only constraints on the radial velocity component. The $d^\tm{RZA}$ distribution function revealed that this fundamental limitation for RZA may be more significant than suggested by just the overall reconstruction rms error. Let us consider the ``viewing angle'' $\gamma$ of a single peculiar velocity observation,
\begin{align}
\label{eq:viewingangle}
\gamma=\tm{acos}\left(\frac{|\bl v \cdot \bl r|}{|\bl v| \cdot |\bl r|}\right)\;\;\;,
\end{align}
which is the angle between the full three-dimensional velocity vector of an object and its observed radial component. At $\gamma=0^\circ$, $v_r$ will have the complete information on an object's peculiar velocity, while at $\gamma=90^\circ$, the object's motion will be completely obscured. The majority of objects in the 30 Mpc/$h$ mock volumes have, in this sense, ``unfavourable'' viewing angles: $\gamma>45^\circ$ for 67\% of the haloes. We can individually compare the reconstruction quality for $\bl \psi^{\tm{WF}}$ for objects depending on their viewing angle to quantify its impact. 

Figure \ref{fig:viewingangle} shows a binning of $d^{\tm{RZA}}$ over $\gamma$. Along with the models in Figure \ref{fig:binnedmodels}, it also includes the C30\_3D mock reconstruction, which has the same datapoints as C30\_00, but with all three components, and is therefore unbiased with respect to $\gamma$. In comparison with this, for the radial velocity reconstructions (coloured bars) there is a tendency of higher $d^\tm{RZA}$ with increasing $\gamma$. This is expected: it is more difficult for the WF to reconstruct the displacement field at a position where the data has a high $\gamma$. It is particularly instructive to compare the 30$^\circ$ -- 45$^\circ$ bin with the 75$^\circ$ -- 90$^\circ$ bin. In both, the unbiased ideal RZA and C30\_3D have similar values, but the radial velocity reconstructions (coloured) show a higher $d^\tm{RZA}$ scatter in the 75$^\circ$ -- 90$^\circ$ bin. This additional scatter is however only at the order of 1 -- 2 Mpc/$h$. This means that the WF yields a reasonable reconstruction for $\bl \psi^{\tm{WF}}$ even for datapoints at unfavourable viewing angles.

We can see the net effect of radial vs. 3D data by directly comparing C30\_3D to C30\_00 (purple). Then, the effect of having only radial data can be decomposed into an additional local error at the most affected positions (high $\gamma$), causing some of the $d^\tm{RZA}$ values to be significantly higher (more skewed $d^\tm{RZA}$ distribution), and an error affecting the reconstruction as a whole by increasing $d^\tm{RZA}$ by 1 -- 1.5 Mpc/$h$ on average. These errors are remarkably small compared to how much information on the full velocity vector is obscured by the radial limitation. Therefore we can state that the WF + RZA procedure performs very well on radial velocity data.  Most importantly, from Figure \ref{fig:viewingangle} it is clear that the radial limitation is not the dominating error source of RZA reconstruction.

%%%%%%%%%%%%%%%%%%%%%%%%%%%%%%%%%%%%%%%%%%%%%%%%%%%%%%%%%%%%%%%%%%%%%%%%%%%
\section{Summary and discussion}
\label{sec:summary}

In this paper, we investigated the Reverse Zeldovich Approximation (RZA), a Lagrangian reconstruction method to trace a galaxy peculiar velocity dataset back in time to early redshifts in the linear regime, after which the data can be used to constrain cosmological initial conditions and to run constrained simulations of the Local Universe. In particular, we tested how the reconstruction quality of the RZA method is affected by observational errors and limitations present in such datasets. For this, we applied the method to a set of mock peculiar velocity catalogues extracted from a cosmological simulation. We investigated the influence of distance measurement errors, data sparseness, limited data volume, different object selection methods, and the effect of being limited to only one of three cartesian components of the galaxy peculiar velocity vector, by varying all these effects in the different mock catalogues. We then used the true initial conditions of the reference simulations to determine how well the cosmological displacements $\bl \psi$ and the initial positions $\bl x_\tm{init}$ can be recovered with RZA. 

With our results we can confirm that RZA significantly improves the reconstruction quality compared to the previous method not using Lagrangian reconstruction. In our sample, RZA reduces the initial position errors from 11 Mpc/$h$ to around 4 -- 5 Mpc/$h$ for a realistic mock data quality. Aside from this, our main conclusion is that the accuracy of the RZA reconstruction is limited by the inherent non-linearity of the velocity field at $z=0$. The effect of this non-linearity is always stronger than
 the combined effect of observational errors even for poor quality datasets. This non-linearity manifests itself in the fact that the displacement field $\bl \psi$ and the peculiar velocity field $\bl v$ diverge from their linear-theory interrelation on a scale of 3.7 Mpc/$h$ on average in our sample, which sets a hard limit for possible Lagrangian reconstruction from peculiar velocities. This limit also highly depends on the local density.

We therefore conclude that observational errors present in peculiar velocity datasets do not present a major obstacle for applying RZA reconstruction. Even an extremely sparse dataset with high observational errors still leads to a good reconstruction of the initial conditions with the median error on the initial positions being $\approx 5$ Mpc/$h$. That said, a significant increase in reconstruction quality can be obtained by increasing the number density of data points, i.e.\ by working with less sparse data. The observational distance errors and the radial-component limitation have a lesser influence. One can compensate very well for the uncertainties they introduce by applying a Wiener Filter reconstruction to the data prior to RZA. We surprisingly find that not knowing two of the three velocity components introduces an error of only $1 - 1.5$ Mpc/$h$ on average.

We also find that a better reconstruction is obtained with a more homogeneous sample of data points, providing a more complete mapping of the volume, and the reconstruction quality worsens if the data points are biased towards the most massive objects and therefore preferentially located in high-density regions. This translates to the statement that observational distance measurement methods selecting galaxies in a more random fashion and not biased towards high-density environments provide ideal input data for RZA. This is the case for peculiar velocities obtained with the Tully-Fisher method, which  selects spiral galaxies on the random basis of their inclination on the sky being grater than 45 degrees, and ignores elliptical galaxies that are strongly biased towards high-density regions. We therefore argue that Tully-Fisher data is best suited for RZA reconstruction compared to peculiar velocity data obtained from other distance measurement methods.

In an upcoming work (Paper III of the series on RZA) we will present constrained simulations run from initial conditions obtained with RZA reconstruction, and analyze their accuracy. Ultimately, the goal of our work is to apply the method directly to the newest observational data to obtain constrained simulations of the Local Universe. With the findings of the analysis presented here, we can expect that the accuracy in which they will reproduce the large-scale structure of the Local Universe will be much higher than that of previous constrained simulations.

%%%%%%%%%%%%%%%%%%%%%%%%%%%%%%%%%%%%%%%%%%%%%%%%%%%%%%

\section*{Acknowledgments}
TD would like to thank R. Brent Tully, Matthias Steinmetz, Francisco-Shu Kitaura, Jochen Klar, Adrian Partl, Steffen Knollmann, Noam I Libeskind, Steffen Hess, Guilhem Lavaux, Saleem Zaroubi, and Alexander Knebe for helpful and stimulating discussions.
YH and SG acknowledge support by DFG  under GO 563/21-1.
YH has been partially supported by the Israel Science Foundation (13/08).
TD and SG acknowledge support by DAAD for the collaboration with H.M. Courtois and R.B. Tully.
We would like to thank the referee of this series of papers for her/his very careful and fast reading of the manuscripts and the many constructive comments which improved the three papers substantially.
%%%%%%%%%%%%%%%%%%%%%%%%%%%%%%%%%%%%%%%%%%%%%%%%%%%%%%

\bibliography{Doumler2012_paper2} \bsp

\begin{thebibliography}{}

\bibitem[\protect\citeauthoryear{{Bardeen}, {Bond}, {Kaiser} \&
  {Szalay}}{{Bardeen} et~al.}{1986}]{Bardeen1986}
{Bardeen} J.~M.,  {Bond} J.~R.,  {Kaiser} N.,    {Szalay} A.~S.,  1986, \apj,
  304, 15

\bibitem[\protect\citeauthoryear{{Bistolas} \& {Hoffman}}{{Bistolas} \&
  {Hoffman}}{1998}]{Bistolas1998}
{Bistolas} V.,  {Hoffman} Y.,  1998, \apj, 492, 439

\bibitem[\protect\citeauthoryear{{Colless}, {Saglia}, {Burstein}, {Davies},
  {McMahan} \& {Wegner}}{{Colless} et~al.}{2001}]{Colless2001}
{Colless} M.,  {Saglia} R.~P.,  {Burstein} D.,  {Davies} R.~L.,  {McMahan}
  R.~K.,    {Wegner} G.,  2001, \mnras, 321, 277

\bibitem[\protect\citeauthoryear{{Courtois}, {Hoffman}, {Tully} \&
  {Gottlober}}{{Courtois} et~al.}{2012}]{Courtois2012}
{Courtois} H.~M.,  {Hoffman} Y.,  {Tully} R.~B.,    {Gottlober} S.,  2012,
  \apj, 744, 43

\bibitem[\protect\citeauthoryear{{Courtois} \& {Tully}}{{Courtois} \&
  {Tully}}{2012}]{Courtois2012arXiv}
{Courtois} H.~M.,  {Tully} R.~B.,  2012, \apj, 749, 174

\bibitem[\protect\citeauthoryear{{Courtois}}{{Courtois}}{2011a}]{Courtois2011a}
{Courtois} H.~M. e.~a.,  2011a, \mnras, 414, 2005

\bibitem[\protect\citeauthoryear{{Courtois}}{{Courtois}}{2011b}]{Courtois2011b}
{Courtois} H.~M. e.~a.,  2011b, \mnras, 415, 1935

\bibitem[\protect\citeauthoryear{{Cuesta}, {Jeltema}, {Zandanel}, {Profumo},
  {Prada}, {Yepes}, {Klypin}, {Hoffman}, {Gottl{\"o}ber}, {Primack},
  {S{\'a}nchez-Conde} \& {Pfrommer}}{{Cuesta} et~al.}{2011}]{Cuesta2011}
{Cuesta} A.~J.,  {Jeltema} T.~E.,  {Zandanel} F.,  {Profumo} S.,  {Prada} F.,
  {Yepes} G.,  {Klypin} A.,  {Hoffman} Y.,  {Gottl{\"o}ber} S.,  {Primack} J.,
  {S{\'a}nchez-Conde} M.~A.,    {Pfrommer} C.,  2011, \apjl, 726, L6

\bibitem[\protect\citeauthoryear{{Djorgovski} \& {Davis}}{{Djorgovski} \&
  {Davis}}{1987}]{Djorgovski1987}
{Djorgovski} S.,  {Davis} M.,  1987, \apj, 313, 59

\bibitem[\protect\citeauthoryear{{Faber} \& {Jackson}}{{Faber} \&
  {Jackson}}{1976}]{Faber1976}
{Faber} S.~M.,  {Jackson} R.~E.,  1976, \apj, 204, 668

\bibitem[\protect\citeauthoryear{{Freedman} \& {Madore}}{{Freedman} \&
  {Madore}}{1990}]{Freedman1990}
{Freedman} W.~L.,  {Madore} B.~F.,  1990, \apj, 365, 186

\bibitem[\protect\citeauthoryear{{Freedman}, {Madore}, {Gibson}, {Ferrarese},
  {Kelson}, {Sakai}, {Mould}, {Kennicutt} Jr., {Ford}, {Graham}, {Huchra},
  {Hughes}, {Illingworth}, {Macri} \& {Stetson}}{{Freedman}
  et~al.}{2001}]{Freedman2001}
{Freedman} W.~L.,  {Madore} B.~F.,  {Gibson} B.~K.,  {Ferrarese} L.,  {Kelson}
  D.~D.,  {Sakai} S.,  {Mould} J.~R.,  {Kennicutt} Jr. R.~C.,  {Ford} H.~C.,
  {Graham} J.~A.,  {Huchra} J.~P.,  {Hughes} S.~M.~G.,  {Illingworth} G.~D.,
  {Macri} L.~M.,    {Stetson} P.~B.,  2001, \apj, 553, 47

\bibitem[\protect\citeauthoryear{{Gottl\"ober}, {Hoffman} \&
  {Yepes}}{{Gottl\"ober} et~al.}{2010}]{Gottloeber2010arXiv}
{Gottl\"ober} S.,  {Hoffman} Y.,    {Yepes} G.,  2010, {ArXiv e-prints}

\bibitem[\protect\citeauthoryear{{Hamana}, {Kayo}, {Yoshida}, {Suto} \&
  {Jing}}{{Hamana} et~al.}{2003}]{Hamana2003}
{Hamana} T.,  {Kayo} I.,  {Yoshida} N.,  {Suto} Y.,    {Jing} Y.~P.,  2003,
  \mnras, 343, 1312

\bibitem[\protect\citeauthoryear{{Hoffman} \& {Ribak}}{{Hoffman} \&
  {Ribak}}{1991}]{Hoffman1991}
{Hoffman} Y.,  {Ribak} E.,  1991, \apjl, 380, L5

\bibitem[\protect\citeauthoryear{{Jha}, {Riess} \& {Kirshner}}{{Jha}
  et~al.}{2007}]{Jha2007}
{Jha} S.,  {Riess} A.~G.,    {Kirshner} R.~P.,  2007, \apj, 659, 122

\bibitem[\protect\citeauthoryear{{Karachentsev}, {Karachentseva}, {Huchtmeier}
  \& {Makarov}}{{Karachentsev} et~al.}{2004}]{Karachentsev2004}
{Karachentsev} I.~D.,  {Karachentseva} V.~E.,  {Huchtmeier} W.~K.,    {Makarov}
  D.~I.,  2004, \aj, 127, 2031

\bibitem[\protect\citeauthoryear{{Klypin}, {Hoffman}, {Kravtsov} \&
  {Gottl{\"o}ber}}{{Klypin} et~al.}{2003}]{Klypin2003}
{Klypin} A.,  {Hoffman} Y.,  {Kravtsov} A.~V.,    {Gottl{\"o}ber} S.,  2003,
  \apj, 596, 19

\bibitem[\protect\citeauthoryear{{Knollmann} \& {Knebe}}{{Knollmann} \&
  {Knebe}}{2009}]{Knollmann2009}
{Knollmann} S.~R.,  {Knebe} A.,  2009, \apjs, 182, 608

\bibitem[\protect\citeauthoryear{{Kolatt}, {Dekel}, {Ganon} \&
  {Willick}}{{Kolatt} et~al.}{1996}]{Kolatt1996}
{Kolatt} T.,  {Dekel} A.,  {Ganon} G.,    {Willick} J.~A.,  1996, \apj, 458,
  419

\bibitem[\protect\citeauthoryear{{Lavaux}}{{Lavaux}}{2010}]{Lavaux2010b}
{Lavaux} G.,  2010, \mnras, 406, 1007

\bibitem[\protect\citeauthoryear{{Mathis}, {Lemson}, {Springel}, {Kauffmann},
  {White}, {Eldar} \& {Dekel}}{{Mathis} et~al.}{2002}]{Mathis2002}
{Mathis} H.,  {Lemson} G.,  {Springel} V.,  {Kauffmann} G.,  {White} S.~D.~M.,
  {Eldar} A.,    {Dekel} A.,  2002, \mnras, 333, 739

\bibitem[\protect\citeauthoryear{{Nusser}, {Branchini} \& {Davis}}{{Nusser}
  et~al.}{2012}]{Nusser2012arXiv}
{Nusser} A.,  {Branchini} E.,    {Davis} M.,  2012, {ArXiv e-prints}

\bibitem[\protect\citeauthoryear{{Rizzi}, {Tully}, {Makarov}, {Makarova},
  {Dolphin}, {Sakai} \& {Shaya}}{{Rizzi} et~al.}{2007}]{Rizzi2007}
{Rizzi} L.,  {Tully} R.~B.,  {Makarov} D.,  {Makarova} L.,  {Dolphin} A.~E.,
  {Sakai} S.,    {Shaya} E.~J.,  2007, \apj, 661, 815

\bibitem[\protect\citeauthoryear{{Shandarin} \& {Zeldovich}}{{Shandarin} \&
  {Zeldovich}}{1989}]{Shandarin1989}
{Shandarin} S.~F.,  {Zeldovich} Y.~B.,  1989, {Reviews of Modern Physics}, 61,
  185

\bibitem[\protect\citeauthoryear{{Sheth} \& {Diaferio}}{{Sheth} \&
  {Diaferio}}{2001}]{Sheth2001b}
{Sheth} R.~K.,  {Diaferio} A.,  2001, \mnras, 322, 901

\bibitem[\protect\citeauthoryear{{Tonry}, {Dressler}, {Blakeslee}, {Ajhar},
  {Fletcher}, {Luppino}, {Metzger} \& {Moore}}{{Tonry}
  et~al.}{2001}]{Tonry2001}
{Tonry} J.~L.,  {Dressler} A.,  {Blakeslee} J.~P.,  {Ajhar} E.~A.,  {Fletcher}
  A.~B.,  {Luppino} G.~A.,  {Metzger} M.~R.,    {Moore} C.~B.,  2001, \apj,
  546, 681

\bibitem[\protect\citeauthoryear{{Tully} \& {Courtois}}{{Tully} \&
  {Courtois}}{2012}]{Tully2012arXiv}
{Tully} R.~B.,  {Courtois} H.~M.,  2012, \apj, 749, 78

\bibitem[\protect\citeauthoryear{{Tully} \& {Fisher}}{{Tully} \&
  {Fisher}}{1977}]{Tully1977}
{Tully} R.~B.,  {Fisher} J.~R.,  1977, \aap, 54, 661

\bibitem[\protect\citeauthoryear{{Tully}, {Shaya}, {Karachentsev}, {Courtois},
  {Kocevski}, {Rizzi} \& {Peel}}{{Tully} et~al.}{2008}]{Tully2008}
{Tully} R.~B.,  {Shaya} E.~J.,  {Karachentsev} I.~D.,  {Courtois} H.~M.,
  {Kocevski} D.~D.,  {Rizzi} L.,    {Peel} A.,  2008, \apj, 676, 184

\bibitem[\protect\citeauthoryear{{van der Wel}, {Bell}, {Holden}, {Skibba} \&
  {Rix}}{{van der Wel} et~al.}{2010}]{Wel2010}
{van der Wel} A.,  {Bell} E.~F.,  {Holden} B.~P.,  {Skibba} R.~A.,    {Rix}
  H.-W.,  2010, \apj, 714, 1779

\bibitem[\protect\citeauthoryear{{Zaroubi}, {Hoffman} \& {Dekel}}{{Zaroubi}
  et~al.}{1999}]{Zaroubi1999}
{Zaroubi} S.,  {Hoffman} Y.,    {Dekel} A.,  1999, \apj, 520, 413

\bibitem[\protect\citeauthoryear{{Zaroubi}, {Hoffman}, {Fisher} \&
  {Lahav}}{{Zaroubi} et~al.}{1995}]{Zaroubi1995}
{Zaroubi} S.,  {Hoffman} Y.,  {Fisher} K.~B.,    {Lahav} O.,  1995, \apj, 449,
  446

\bibitem[\protect\citeauthoryear{{Zeldovich}}{{Zeldovich}}{1970}]{Zeldovich1970}
{Zeldovich} Y.~B.,  1970, \aap, 5, 84

\end{thebibliography}

\label{lastpage}

\newpage

\end{document}